\documentclass{article}

\usepackage[final]{neurips_2021}

\usepackage[utf8]{inputenc} 
\usepackage[T1]{fontenc}    
\usepackage{url}            
\usepackage{booktabs}       
\usepackage{amsfonts}       
\usepackage{nicefrac}       
\usepackage{microtype}      
\usepackage{xcolor}         

\usepackage{graphicx}
\usepackage{dsfont}
\usepackage{amsfonts}
\usepackage{amsmath}
\usepackage{mathtools}
\usepackage{supertabular}
\usepackage{mathrsfs,amssymb}
\usepackage{color}
\usepackage{listings}
\usepackage{algorithm}
\usepackage[noend]{algpseudocode}
\usepackage{tikz}
\usetikzlibrary{arrows}
\usepackage[T1]{fontenc}
\usepackage[Symbol]{upgreek}
\usepackage{arydshln}
\usepackage{xcolor}
\usepackage{soul}
\usepackage{hyphenat}

\usepackage{amsthm}

\usepackage{algorithm}
\usepackage[noend]{algpseudocode}
\usepackage{tikz}
\usepackage{mathtools}

\usepackage{mathtools}

\usepackage{multirow}
\usepackage{multicol}

\usepackage{subfigure}

\usepackage[utf8]{inputenc}
\usepackage{pgfplots}
\DeclareUnicodeCharacter{2212}{−}
\usepgfplotslibrary{groupplots,dateplot}
\usetikzlibrary{patterns,shapes.arrows}
\pgfplotsset{compat=newest}
\pgfplotsset{scaled y ticks=false}

\usepackage{caption}
\usepackage{multirow}
\usepackage{multicol}

\def\munderbar#1{\underline{\sbox\tw@{$#1$}\dp\tw@\z@\box\tw@}}
\usepackage{csquotes}

\title{On Parameter Estimation in Unobserved Components Models subject to Linear Inequality Constraints}

\author{%
  Abhishek K.~Umrawal 
    \\
  Purdue University \\
  West Lafayette, 47906 \\
  \texttt{aumrawal@purdue.edu} \\
  \And
  Joshua C. C.~Chan \\
  Purdue University \\
  West Lafayette, 47906 \\
  \texttt{chan196@purdue.edu} \\
}

\begin{document}

\maketitle
\begin{abstract}
We propose a new \textit{quadratic programming-based} method of approximating a nonstandard density using a multivariate Gaussian density. Such nonstandard densities usually arise while developing posterior samplers for unobserved components models involving inequality constraints on the parameters. For instance, \cite{chan2016bounded} provided a new model of trend inflation with linear inequality constraints on the stochastic trend. We implemented the proposed quadratic programming-based method for this model and compared it to the existing approximation. We observed that the proposed method works as well as the existing approximation in terms of the final trend estimates while achieving gains in terms of sample efficiency.
\end{abstract}
\section{Introduction}
Consider the following unobserved components model.
\begin{align*}
    y_t &= \tau_t + \epsilon_t^y, \ \ \epsilon_t^y \sim \mathcal {N}(0,\sigma^2), \\
    \tau_t &= \tau_{t-1} + \epsilon_t^h \ \ \epsilon_t^h \sim \mathcal {N}(0,\omega^2), 
\end{align*}
where the initial condition $\tau_0$ is treated as a parameter.

Suppose we wish to restrict the trend $\tau_t$ to be in the interval $[a_t,b_t], t=1,...,T$ where $a_t$ and $b_t$ are constants. Linear inequality constraints like these occur quite naturally in many economic applications. For instance, \cite{chan2016bounded} discussed such situations in the context of modeling trend inflation. The problem of interest is to efficiently estimate the parameters involved in an unobserved components model under such restrictions. By efficiency, we mean the run time and the sample efficiency.

\subsection{Motivation}
In the presence of the aforementioned inequality constraints, the conditional distribution of $\boldsymbol{\tau}$ given the other parameters, i.e. $p(\boldsymbol{\tau} | \mathbf{y}, \sigma^2, \omega^2, \tau_0)$ is nonstandard and thus conventional methods of inference in state-space models cannot be used. \cite{chan2012estimation} suggest an approach for estimation in such non-linear state-space models where the key element is to approximate $p(\boldsymbol{\tau} | \mathbf{y}, \sigma^2, \omega^2, \tau_0)$ using a multivariate Gaussian density. The Gaussian approximation is then used as a proposal density for an acceptance-rejection Metropolis-Hastings algorithm (\cite{chan2016bounded}).

We propose a new quadratic programming-based approach to approximating the nonstandard density using a Gaussian density. In general, the idea is to efficiently solve the following quadratic programming problem. 
\begin{align*}
    \max_{\boldsymbol{\tau} } & \ -\frac{1}{2} (\boldsymbol{\tau} -\mathbf{m})'\mathbf{C}(\boldsymbol{\tau} -\mathbf{m}),\\
    \text{subject to } & \ \mathbf{a} \le \mathbf{A} \boldsymbol{\tau} \le \mathbf{b}.
\end{align*}

The above quadratic program maximizes the kernel of a multivariate Gaussian density for $\boldsymbol{\tau}$ for general linear inequality constraints. In general, we can say the following about the time complexity of solving a quadratic program like the one mentioned above.
\begin{enumerate}
    \item When $\mathbf{C}$ is a positive-definite matrix, the ellipsoid method solves the problem in (weakly) polynomial time (\cite{kozlov1980polynomial}). This is our case.
    \item When $\mathbf{C}$ is indefinite, then the problem is NP-hard (\cite{sahni1974computationally}).
\end{enumerate}

As our problem has more structure like $\mathbf{A} = \mathbf{I}_T$. We may be able to further improve on the empirical run time if not the asymptotic polynomial time complexity.

\subsection{Investigating Run-Time of the Quadratic Program}

We know the following.
$$p(\boldsymbol{\tau} | \mathbf{y}, \sigma^2, \omega^2, \tau_0) \sim \mathcal {N}(\hat{\boldsymbol \tau},\mathbf{K}_{\boldsymbol\tau}^{-1}),$$  

where $\mathbf{K}_{\boldsymbol\tau} = \left(\frac{1}{\omega^2} + \frac{1}{\sigma^2} \right)\mathbf{H}'\mathbf{H}$, $\hat{\boldsymbol \tau}  = \mathbf{K}_{\boldsymbol\tau}^{-1} \left(\frac{\tau_0}{\omega^2}\mathbf{H}'\mathbf{H} + \frac{1}{\sigma^2}\mathbf{H}'\mathbf{H}\mathbf{y}\right)$, and $\mathbf{H}$ is given as follows.

\begin{equation*}
    \mathbf{H} = \begin{pmatrix*}[l]
    \phantom{-}1 & 0 & ... & \phantom{-}0 & 0 \\
    -1 & 1 & ... & \phantom{-}0 & 0\\
     & &... & & \\
    \phantom{-}0 & 0 & ... & -1 & 1\\
    \end{pmatrix*}.
\end{equation*}

We set $\mathbf{m}=\hat{\boldsymbol \tau}$, $\mathbf{C}=\mathbf{K}_{\boldsymbol\tau}$, and $\mathbf{A} = \mathbf{I}_T$. We use \texttt{quadprogram} available in MATLAB to solve the following quadratic programming problem. 
\begin{align*}
    \max_{\boldsymbol{\tau} } & \ -\frac{1}{2} (\boldsymbol{\tau} -\mathbf{m})'\mathbf{C}(\boldsymbol{\tau} -\mathbf{m}),\\
    \text{subject to } & \ \mathbf{a} \le \boldsymbol{\tau} \le \mathbf{b}.
\end{align*}

We can write it equivalently as follows.
\begin{align*}
    \min_{\boldsymbol{\tau} } & \ \frac{1}{2} \boldsymbol{\tau}'\mathbf{C}\boldsymbol{\tau} - \mathbf{m}'\mathbf{C}\boldsymbol{\tau} + \frac{1}{2}\mathbf{m}'\mathbf{C}\mathbf{m}, \\
    \Leftrightarrow \min_{\boldsymbol{\tau} } & \ \frac{1}{2} \boldsymbol{\tau}'\mathbf{C}\boldsymbol{\tau} + (-\mathbf{C}'\mathbf{m})'\boldsymbol{\tau},\\
    \text{subject to } & \ \mathbf{a} \le \boldsymbol{\tau} \le \mathbf{b}.
\end{align*}

We use the US quarterly CPI data from 1947Q1 to 2015Q4. Specifically, given the quarterly CPI figures $z_t$, we compute $y_t = 400(\log z_t - \log z_{t-1})$ and use it as the inflation rate. We take subsets of $\mathbf{m}$ and $\mathbf{C}$ to investigate the run times for different values of $T$. 

The run times in solving the above quadratic programming problem for different values of $T$ are reported in Table 1.

\begin{table}[ht!]
\caption{Elapsed times in solving the quadratic program}
\centering
\begin{tabular}{c c}
\hline \hline
$T$ & Elapsed Time (in seconds)\\
\hline
50 & 0.0058\\
100 & 0.0076\\
150 & 0.0140\\
200 & 0.0149\\
250 & 0.0172\\
300 & 0.0185\\
350 & 0.0208\\
400 & 0.0264\\
450 & 0.0241\\
500 & 0.0304\\
550 & 0.0309\\
\hline
\end{tabular}
\end{table}

\textbf{Comment.} We observed that the empirical run times are quite small. This suggests that the idea is promising and we  should explore it further. The solution to the discussed quadratic program is bound to lie within the constraints. Hence, for Bayesian computations, it is a better candidate for the mean of the distribution which the parameter is getting generated from. The idea is to improve the estimation and avoid any approximations due to the linear constraints and also expect that the sample efficiency will improve. The estimation performed using this idea will be termed \texttt{QuadProg} henceforth.
\section{\texttt{QuadProg} for the Trend Inflation Model discussed in \cite{chan2016bounded}}

\subsection{Model Description}
\cite{chan2016bounded} discuss a new model of trend inflation with inequality constraints. The model is discussed in great detail is the original paper. Let $\pi_t$ is an observed measure of inflation at time $t$, then the measurement and state equations of the model are given as follows.
\begin{align*}
    \pi_t - \tau_t &= \rho_t (\tau_{t-1} - \tau_{t-1}) + \epsilon_t \exp{\left(\frac{h_t}{2}\right)},\\
    \tau_t &= \tau_{t-1} + \epsilon_t^\tau,\\
    \rho_t &= \rho_{t-1} + \epsilon_t^\rho,\\
    h_t &= h_{t-1} + \epsilon_t^h,
\end{align*}

where $\epsilon_t \sim \mathcal{N}(0,1)$ and $\epsilon_t^h \sim \mathcal{N}(0,\sigma^2)$. The linear inequality constraints on the behavior of $\tau_t$ an $\rho_t$ are imposed by considering the following.
\begin{align*}
    \epsilon_t^\tau &\sim \mathcal{TN}(a_\tau-\tau_{t-1},b_\tau-\tau_{t-1};0,\sigma_\tau^2),\\
    \epsilon_t^\rho &\sim \mathcal{TN}(a_\rho-\rho_{t-1},b_\rho-\rho_{t-1};0,\sigma_\rho^2),
\end{align*}

where $\mathcal{TN}(a,b;\mu,\sigma^2)$ denotes the Gaussian distribution with mean $\mu$ and variance $\sigma^2$ truncated to the interval $(a, b)$.

Furthermore, the model considers the following.
\begin{align*}
    \tau_1 &\sim \mathcal{TN}(a_\tau,b_\tau;\tau_0,\omega_\tau^2),\\
    \rho_1 &\sim \mathcal{TN}(0,1;\rho_0,\omega_\rho^2),\\
    h_1 &\sim \mathcal{N}(h_0,\omega_h^2),
\end{align*}

where $a_\tau,b_\tau,\tau_0,\omega_\tau^2,\rho_0,\omega_\rho^2,h_0$ and $\omega_h^2$ are known constants. 

Following are the priors for the unknown parameters.
\begin{align*}
    \sigma_\tau^2 &\sim \mathcal{IG}(\nu_\tau,S_\tau),\\
    \sigma_\rho^2 &\sim \mathcal{IG}(\nu_\rho,S_\rho),\\
    \sigma_h^2 &\sim \mathcal{IG}(\nu_h,S_h).
\end{align*}

\subsection{The Proposed Modification}

\cite{chan2016bounded} discusses given an initial value $y_0$ of the CPI inflation rate, use the measurement equation to generate the CPI inflation rate data, $\boldsymbol{y}$, and develop a Markov chain Monte Carlo (MCMC) algorithm for generating from the posterior densities for the above model. Due to the presence of the inequality constraints, $p(\boldsymbol{\tau} | \mathbf{y}, \boldsymbol{\rho}, \boldsymbol{h},\boldsymbol{\theta})$ and $p(\boldsymbol{\rho} | \mathbf{y}, \boldsymbol{\tau}, \boldsymbol{h},\boldsymbol{\theta})$ are nonstandard and thus conventional methods of inference in state-space models cannot be used. Instead, they use an approach developed in \cite{chan2012estimation} for posterior sampling in nonlinear state-space models. An essential element of this algorithm is a Gaussian approximation to the conditional density $p(\boldsymbol{\tau} | \mathbf{y}, \boldsymbol{\rho}, \boldsymbol{h},\boldsymbol{\theta})$. This uses a precision-based algorithm that builds upon results derived for the linear Gaussian state-space model by \cite{chan2009efficient}. The Gaussian approximation is then used as a proposal density for an ARMH step. They also use this algorithm to draw from $p(\boldsymbol{\rho} | \mathbf{y}, \boldsymbol{\tau}, \boldsymbol{h},\boldsymbol{\theta})$ and $p(\boldsymbol{h} | \mathbf{y}, \boldsymbol{\tau}, \boldsymbol{\rho},\boldsymbol{\theta})$.


We use the quadratic-programming-based approximation of the Gaussian density (\texttt{QuadProg}) as proposed and discussed in this report for modifying the posterior sampling procedure in \cite{chan2016bounded} for $\boldsymbol{\tau}$ and $\boldsymbol{\rho}$ as follows.

\begin{quote}
    \textbf{Instead of generating from the proposal density of the form $\mathcal{N}(\hat{\boldsymbol\kappa}, D_{\boldsymbol\kappa})$, we can use the quadratic program solver to obtain $\hat{\boldsymbol\kappa}$ and then draw a candidate to be accepted or rejected via an acceptance-rejection Metropolis-Hasting (ARMH) step.}
\end{quote}
\section{Empirical Results}
We used the US quarterly CPI data from 1947Q1 to 2011Q3. Specifically, given the quarterly CPI figures $z_t$, we computed $y_t = 400(\log z_t - \log z_{t-1})$ and used it as the inflation rate. As suggested and discussed in \cite{chan2016bounded}, we set the hyperparameters as follows. 

\begin{enumerate}
    \item $\tau_0 = \rho_0 = h_0 = 0$.
    \item $\omega_\tau^2 = \omega_h^2 = 5$ and $\omega_\rho^2 = 1$.
    \item $\nu_\tau = \nu_\rho = \nu_h = 10$.
    \item $S_\tau = 0.18$, $S_\rho = 0.009$ and $S_h = 0.45$.
\end{enumerate}

We used 10,000 Monte-Carlo simulations with 1000 as the burn-in time. 

We used the following two estimation methods.

\begin{enumerate}
    \item \texttt{AR-Trend-Bound}: Same as discussed in \cite{chan2016bounded}.
    \item \texttt{Quad-Prog}: Estimated method discussed in \cite{chan2016bounded} with the modification proposed in this report.
\end{enumerate}

We used \textsc{Matlab} for the purpose of implementation.

\subsection{Trend Estimation}

We estimated the trend for the US quarterly CPI data using \texttt{AR-Trend-Bound} and \texttt{Quad-Prog} methods.

\begin{figure}
    \centering
    \includegraphics[width=13cm,height=6.5cm]{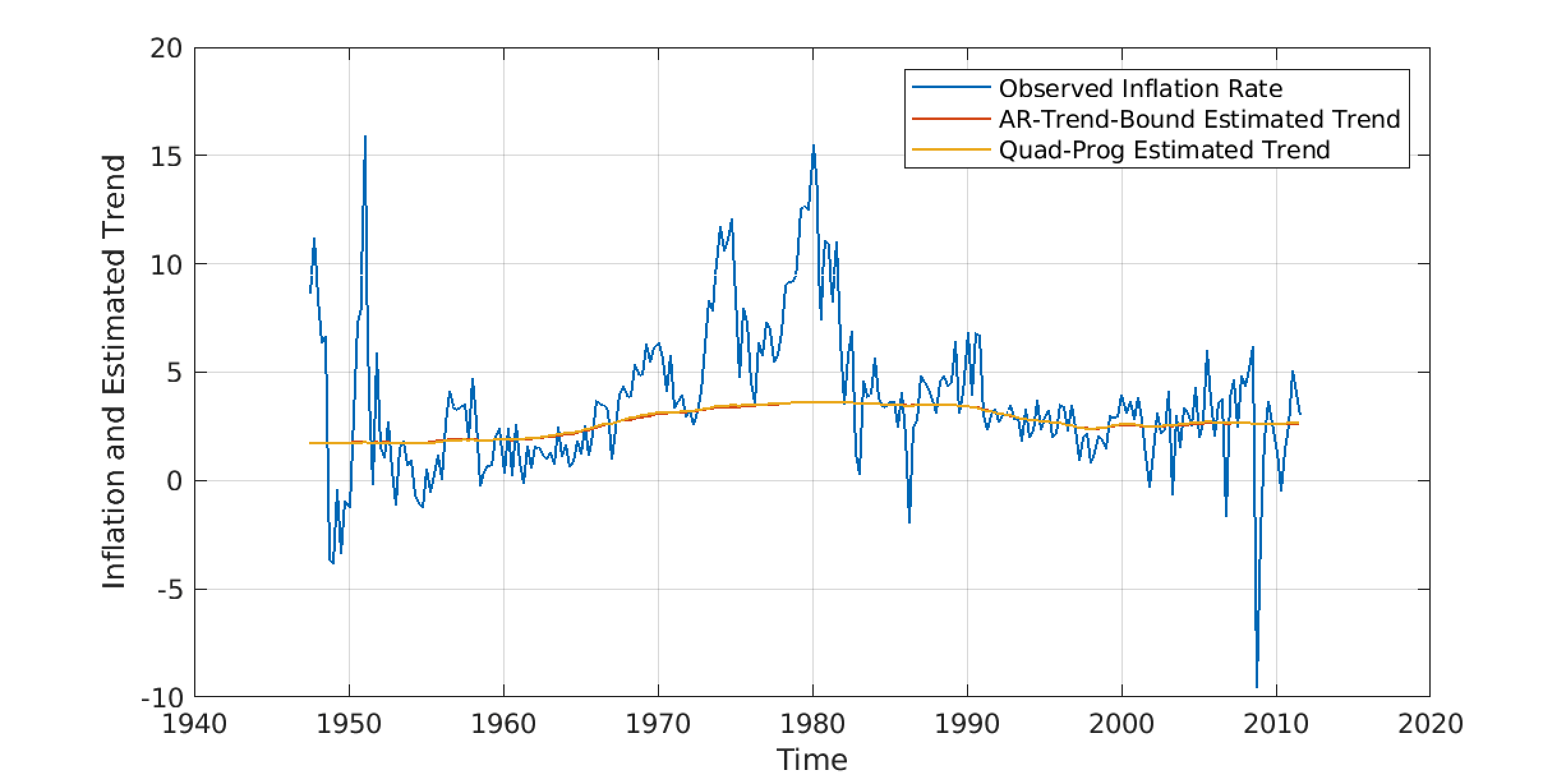}
    \caption{Comparison of trend estimates}
    \label{fig:AR_trend_bound_estimates}
\end{figure}

Figure \ref{fig:AR_trend_bound_estimates} shows the observed rate of inflation, trend estimated using \texttt{AR-Trend-Bound}, and trend estimated using \texttt{Quad-Prog}. We observe that the trend estimates using the two different methods are quite close to each other confirming that the the proposed modification indeed does good estimation.  

\subsection{Sample Efficiency}
One of the limitations of Monte Carlo methods is that different draws are seemingly/pseudo-random, i.e. there may exist a significant autocorrelation in the draws generated over different time points. We are interested in understanding how does the proposed method compares to the existing approach in terms of presence of such autocorrelation. 
A common diagnostic of sample efficiency is the inefficiency factor, defined as follows.
\begin{align*}
    1 + \sum_{l=1}^L \phi_l,
\end{align*}
where $\phi_l$ is the sample autocorrelation at lag length $l$ and $L$ is chosen large enough so that the autocorrelation tapers off. Note that purely random/independent draws would give an inefficiency factor of 1. Inefficiency factors indicate how many extra draws need to be taken to give results equivalent to independent draws. For instance, if we take 5000 draws of a parameter and find an inefficiency factor of 10, then these draws are equivalent to 500 independent draws from a given distribution.

For any given parameter vector of length $T$, there are $T$ inefficiency factors (one for each $t$). As $\boldsymbol{\tau}, \boldsymbol{\rho}$ and $\boldsymbol{h}$ each is of length $T$, we have $3T$ total number of inefficiency factors corresponding each approach of estimation. We used $L=100$ to calculate the inefficiency factors. For reporting, we construct a box plot of $T$ inefficiency factors for each parameter. We compare the box plots for each parameter across different methods of estimation.

\begin{figure}
    \centering
    \includegraphics[width=8cm,height=5.4cm]{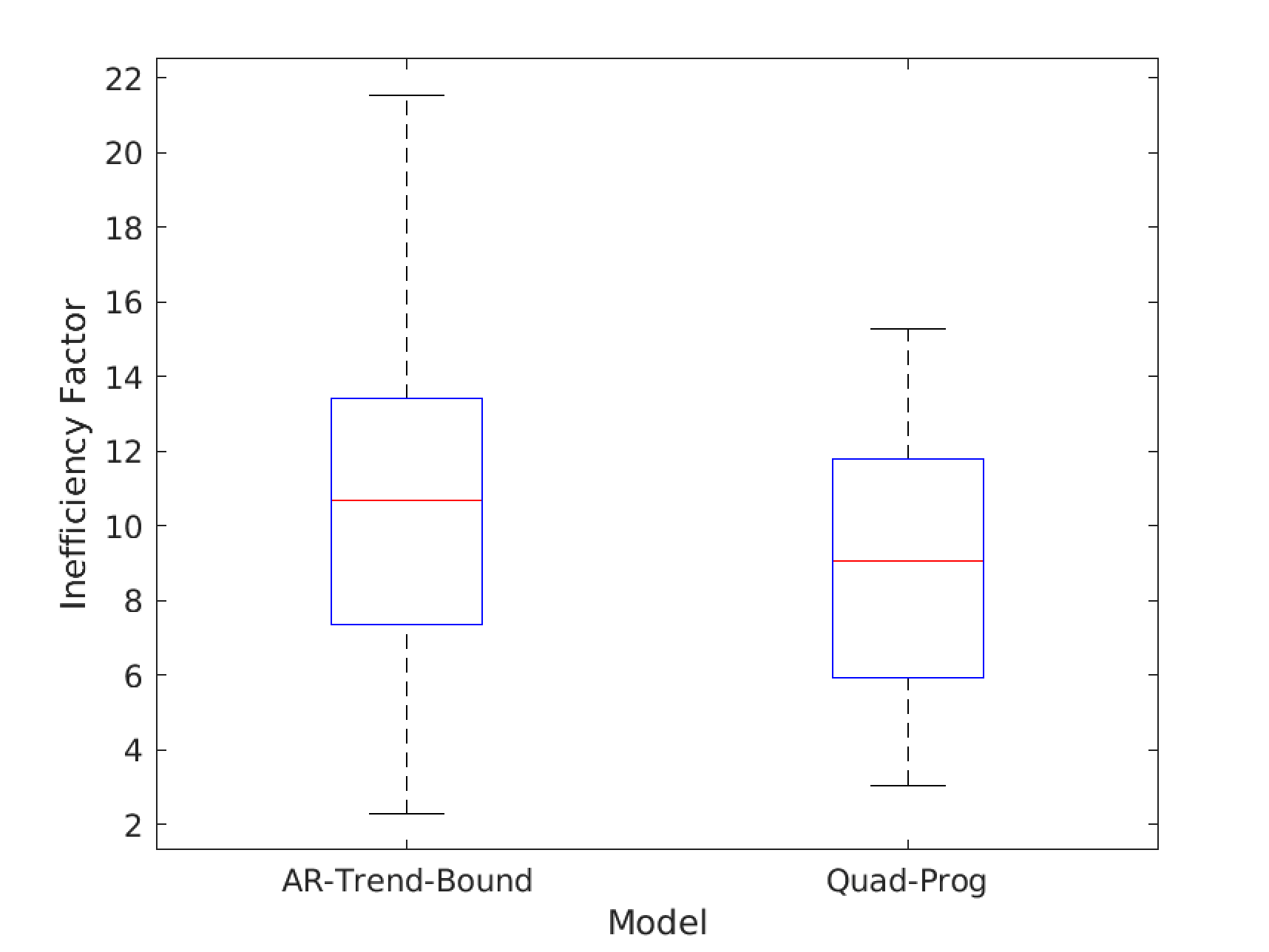}
    \caption{Comparison of inefficiency factors for $\boldsymbol{\tau}$.}
    \label{fig:AR_trend_bound_ineff_tau}
\end{figure}

Figure \ref{fig:AR_trend_bound_ineff_tau} shows the box plots of inefficiency factors for sampling $\boldsymbol{\tau}$ using \texttt{AR-Trend-Bound} and \texttt{Quad-Prog} methods. We observe that \texttt{Quad-Prog} leads to a lower average and maximum sample inefficiency as compared to \texttt{AR-Trend-Bound}.

\begin{figure}
    \centering
    \includegraphics[width=8cm,height=5.4cm]{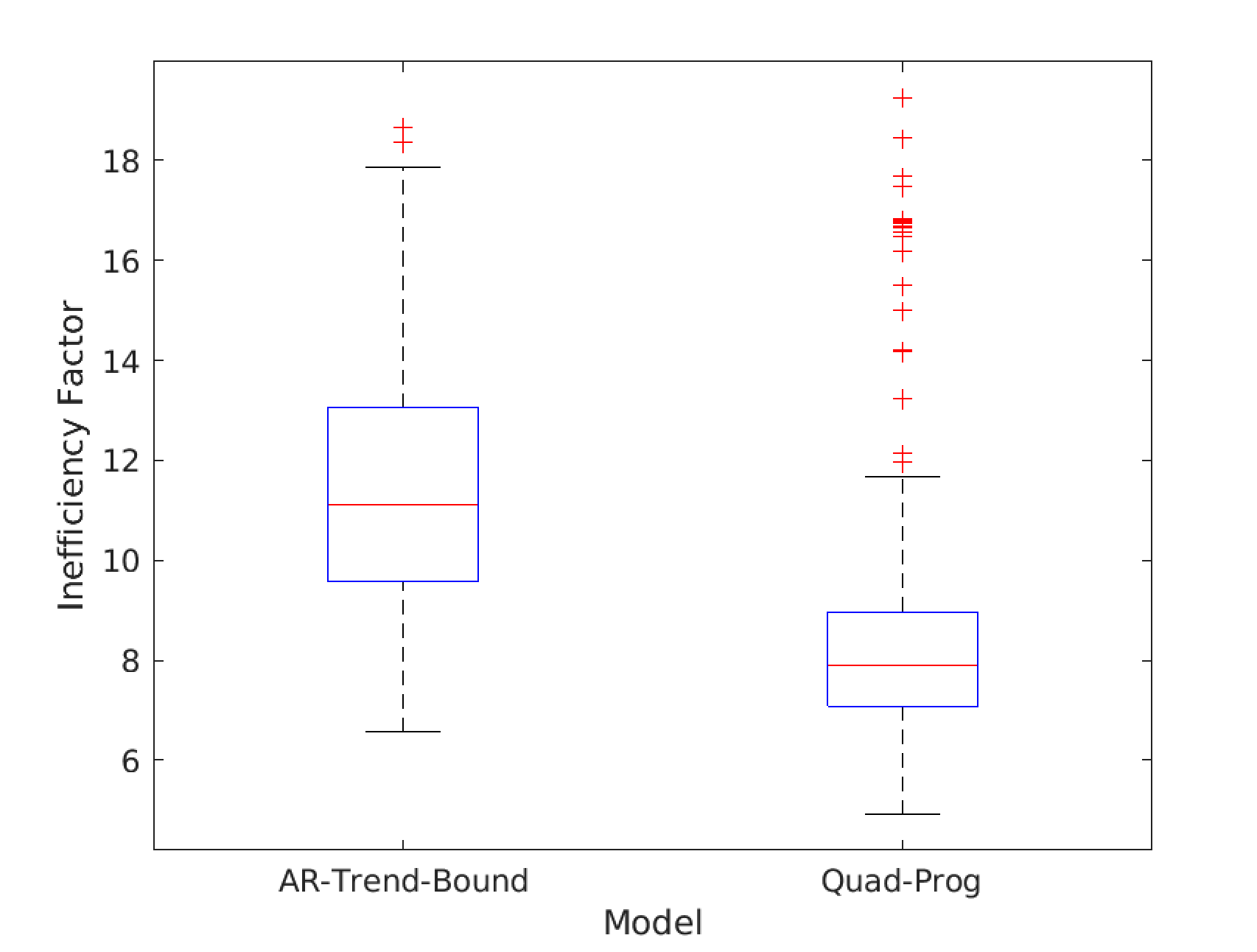}
    \caption{Comparison of inefficiency factors for $\boldsymbol{\rho}$}
    \label{fig:AR_trend_bound_ineff_rho}
\end{figure}

Figure \ref{fig:AR_trend_bound_ineff_rho} shows the box plots of inefficiency factors for sampling $\boldsymbol{\rho}$ using \texttt{AR-Trend-Bound} and \texttt{Quad-Prog} methods. We observe that \texttt{Quad-Prog} leads to a lower average and maximum sample inefficiency as compared to \texttt{AR-Trend-Bound}.

\begin{figure}
    \centering
    \includegraphics[width=8cm,height=5.4cm]{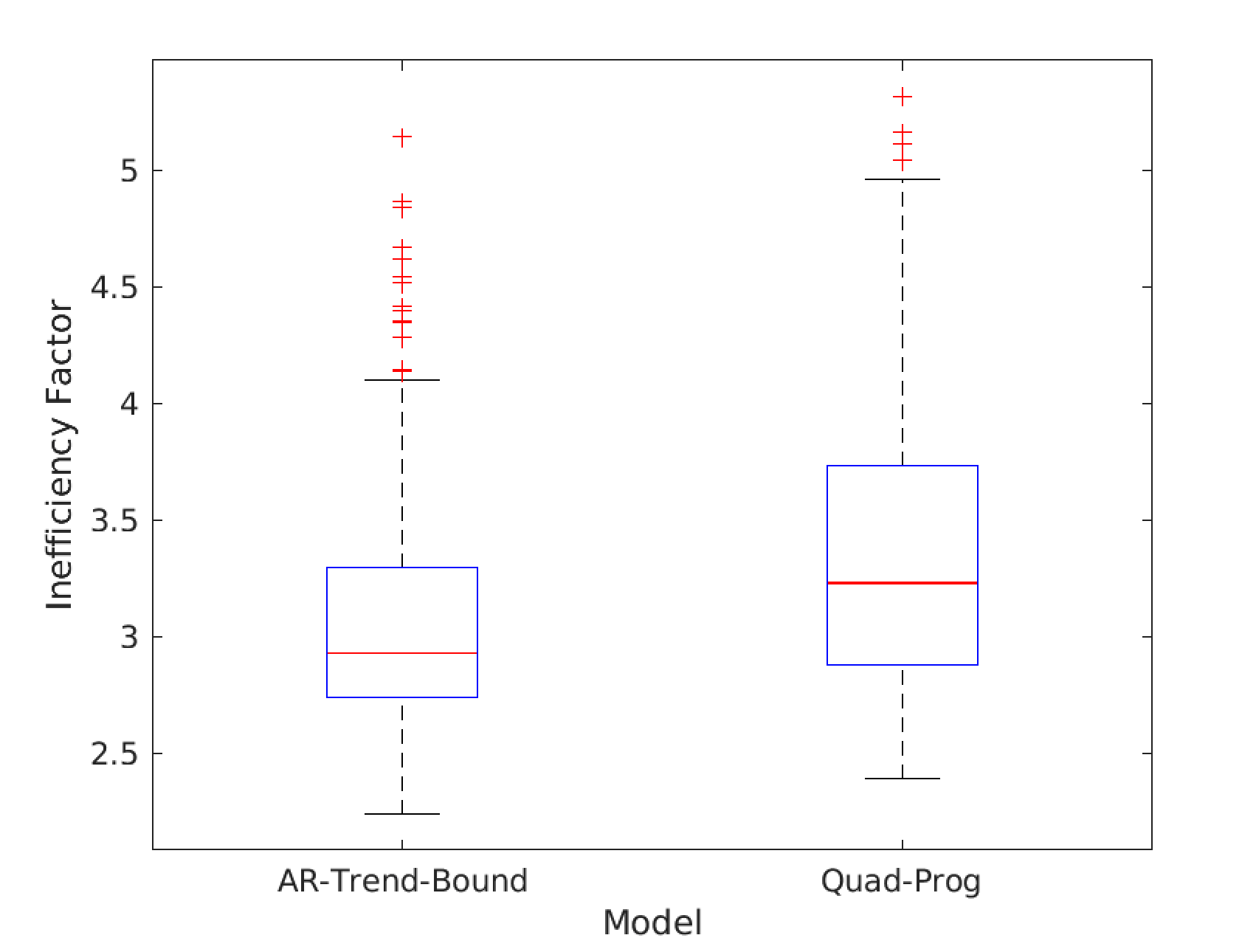}
    \caption{Comparison of inefficiency factors for $\boldsymbol{h}$}
    \label{fig:AR_trend_bound_ineff_h}
\end{figure}

Figure \ref{fig:AR_trend_bound_ineff_h} shows the box plots of inefficiency factors for sampling $\boldsymbol{h}$ using \texttt{AR-Trend-Bound} and \texttt{Quad-Prog} methods. We observe that \texttt{Quad-Prog} \texttt{AR-Trend-Bound} perform very similar in terms of average  and maximum sample efficiency.

\bibliographystyle{apalike}
\bibliography{refs.bib}

\clearpage
\appendix
\section{Implementing \texttt{QuadProg} for Trend Estimation in a Simple Unobserved Components Model (UC-AR)}

As an initial test for the proof-of-concept, we tried the same idea of quadratic programming-based approximation of a multivariate Gaussian density for a relatively simple unobserved components model discussed as follows.

Let $y_t$ be some observed measure at time $t$, then the measurement and state equations of the model are given as follows.
\begin{align*}
    y_t &= \tau_t + \epsilon_t,\\
    \tau_t &= \tau_{t-1} + \eta_t,\\
    \epsilon_t &= \rho \epsilon_{t-1} + u_t,
\end{align*}

where $\eta_t \sim \mathcal{N}(0,\omega^2)$, $\epsilon_0 = 0$, $|\rho|<1$, $u_t \sim \mathcal{N}(0,\sigma^2)$ and $\tau_0$ is an unknown parameter. 

Following are the priors for the unknown parameters.
\begin{align*}
    \tau_0 &\sim \mathcal{N}(a_0,b_0),\\
    \rho &\sim \mathcal{U}(-1,1),\\
    \sigma^2 &\sim \mathcal{IG}(\nu_{\sigma^2},S_{\sigma^2}),\\
    \omega^2 &\sim \mathcal{IG}(\nu_{\omega^2},S_{\omega^2}).
\end{align*}

For the above model, we have an explicit and no approximation posterior sampler, i.e. approximation using a multivariate Gaussian density is not required. However, just to make sure that the proposed idea works well in terms of estimation and sample efficiency, we implemented \texttt{Quad-Prog} for this model.

\subsection{Empirical Results}

We used the US quarterly CPI data from 1947Q1 to 2015Q4. Specifically, given the quarterly CPI figures $z_t$, we computed $y_t = 400(\log z_t - \log z_{t-1})$ and used it as the inflation rate. As suggested and discussed in \cite{chan2016bounded}, we set the hyperparameters as follows. 

\begin{enumerate}
    \item $a_0 = 5$ and $b_0 = 100$.
    \item $\nu_{\sigma^2} = 3$ and $S_{\sigma^2} = \nu_{\sigma^2}-1$
    \item $\nu_{\omega^2} = 3$ and $S_{\omega^2} = .25^2(\nu_{\omega^2}-1)$
\end{enumerate}

We used 10,000 Monte-Carlo simulations with 1000 as the burn-in time. 

We used the following two estimation methods.

\begin{enumerate}
    \item \texttt{UC-AR}: The explicit and no approximation method.
    \item \texttt{Quad-Prog}: Modifying the explicit method using the idea proposed in this report.
\end{enumerate}

We used \textsc{Matlab} for the purpose of implementation.

\subsubsection{Trend Estimation}

We estimated the trend for the US quarterly CPI data using \texttt{UC-AR} and \texttt{Quad-Prog} methods.

\begin{figure}
    \centering
    \includegraphics[width=13cm,height=6.5cm]{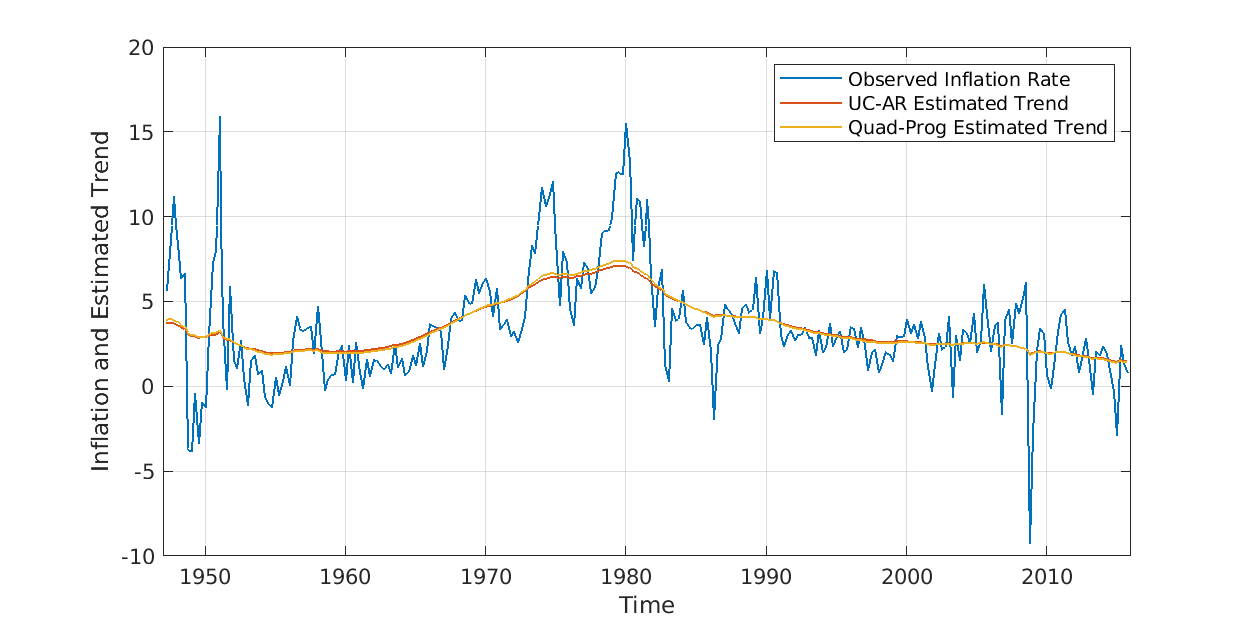}
    \caption{Comparison of trend estimates}
    \label{fig:UC_AR_estimates}
\end{figure}

Figure \ref{fig:UC_AR_estimates} shows the observed rate of inflation, trend estimated using \texttt{UC-AR}, and trend estimated using \texttt{Quad-Prog}. We observe that the trend estimates using the two different methods are quite close to each other confirming that the proposed modification indeed does good estimation.  

\subsubsection{Sample Efficiency}

\begin{figure}
    \centering
    \includegraphics[width=10cm,height=6.5cm]{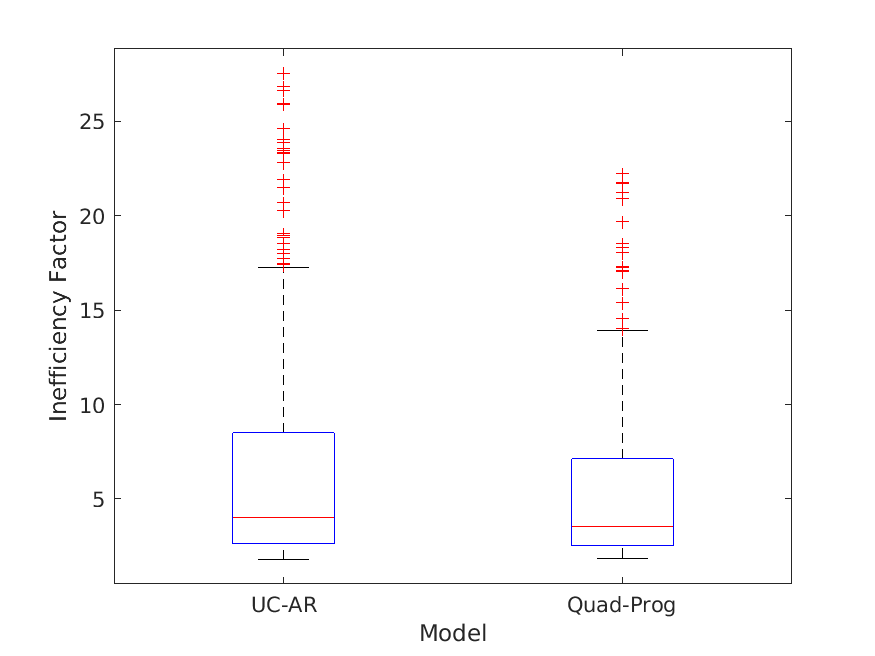}
    \caption{Comparison of inefficiency factors for $\boldsymbol{\tau}$}
    \label{fig:UC_AR_ineff}
\end{figure}

Figure \ref{fig:UC_AR_ineff} shows the box plots of inefficiency factors for sampling $\boldsymbol{\tau}$ using \texttt{AR-Trend-Bound} and \texttt{Quad-Prog} methods. We observe that \texttt{Quad-Prog} and \texttt{UC-AR} perform  similarly in terms of average sample efficiency. However, \texttt{Quad-Prog} leads to a slightly lower maximum sample inefficiency as compared to \texttt{UC-AR} which is most likely attributed to chance.

\end{document}